# Electronic band structure and inter-atomic bonding in layered 1111-like Th-based pnictide oxides ThCuPO, ThCuAsO, ThAgPO, and ThAgAsO from first principles calculations

V.V. Bannikov, I.R. Shein*, A.L. Ivanovskii

*Institute of Solid State Chemistry, Ural Branch of the Russian Academy of Sciences, 620990, Ekaterinburg, Russia*

**A B S T R A C T**

First-principles FLAPW-GGA band structure calculations were employed to examine the structural, electronic properties and the chemical bonding picture for four ZrCuSiAs-like Th-based quaternary pnictide oxides ThCuPO, ThCuAsO, ThAgPO, and ThAgAsO. These compounds were found to be semimetals and may be viewed as "intermediate" systems between two main isostructural groups of superconducting and semiconducting 1111 phases. The Th 5$f$ states participate actively in the formation of valence bands and the Th 5$f$ states for Th$MPn$O phases are itinerant and partially occupied. We found also that the bonding picture in Th$MPn$O phases can be classified as a high-anisotropic mixture of ionic and covalent contributions: inside [Th$_2$O$_2$] and [$M_2Pn_2$] blocks, mixed covalent-ionic bonds take place, whereas between the adjacent [Th$_2$O$_2$]/[$M_2Pn_2$] blocks, ionic bonds emerge owing to [Th$_2$O$_2$] → [$M_2Pn_2$] charge transfer.

*PACS* : 71.18.+y, 71.15.Mb, 74.25.Jb

*Keywords:* ThCuPO; ThCuAsO; ThAgPO; ThAgAsO; Structural, Electronic properties, Inter-atomic bonding, first principles calculations

* Corresponding author.
*E-mail address:* shein@ihim.uran.ru (Igor R. Shein).

## 1. Introduction

The recent discovery [1] of superconductivity in the layered Fe-based pnictide oxides with an unconventional pairing mechanism near the spin-density wave order aroused tremendous interest and in the three last years inspired a great research activity in the field of condensed matter physics and materials science, reviews [2-7].

One of the most extensively studied groups of these materials consists of the so-called 1111 phases, namely, quaternary pnictide oxides *Ln*Fe*Pn*O, where *Ln* are early rare earth metals such as La, Ce, Sm, Dy, Gd *etc.*, and *Pn* are pnictogens. These *Ln*Fe*Pn*O phases adopt a layered ZrCuSiAs-type structure, where $[Ln_2O_2]^{\delta+}$ blocks are sandwiched between $[Fe_2Pn_2]^{\delta-}$ blocks. Let us note that: (i). among more than 70 synthesized ZrCuSiAs-type 1111 layered phases [8], only a few (basically, Fe-containing) were involved till now in the search of superconducting materials, and their electronic properties were examined; (ii). non-doped phases *Ln*Fe*Pn*O are located on the border of magnetic instability and commonly exhibit temperature-dependent structural and magnetic phase transitions with the formation of antiferromagnetic spin ordering, and (iii). superconductivity emerges as a result of hole or electron doping of the parent compounds, see [1-7].

The atomic substitutions inside building blocks of 1111 phases exert a profound influence on their properties, and, in particular, on their superconductivity. Therefore chemical substitution (doping effects) is one of the main strategies for improvement of the properties of these systems. A large number of investigations of the doping effects on the above ZrCuSiAs-like *Ln*Fe*Pn*O phases were performed today, and the most extensively studied situations are $d^{n<10}$ metals substitutions at the Fe site, P substitutions at the As site, and F substitutions at the oxygen site [2-7].

Besides, a set of doped 1111 superconductors were obtained recently using quite an unusual type of substitutions: $Th^{4+} \rightarrow Ln^{3+}$. So, superconductivity through Th substitution at the lanthanide site was found for GdFeAsO ($T_C \sim$ 56K [9]), NdFeAsO ($T_C \sim$ 38K [10]), LaFeAsO ($T_C \sim$ 30K [11]), and SmFeAsO ($T_C \sim$ 50K [12]). However, as distinct from other *sp* and *d* impurities, the influence of Th on the electronic structure of 1111 phases remains actually unstudied.

Besides the above thorium-doped 1111 phases, a set of "pure" ZrCuSiAs-like Th-based pnictide oxides Th*MPn*O, where *M* are Cu or Ag and *Pn* are P or As, were synthesized, see [8].

In this paper, in order to get a insight into the electronic properties and the peculiarities of inter-atomic bonding in thorium-containing 1111 materials, a first-principles study of four thorium-based pnictide oxides: synthesized ThCuPO, ThCuAsO, ThAgPO [8] and hypothetical ThAgAsO was performed. This choice allows us to compare the above properties of these related phases as a function of: (i) the pnictogen type (P *versus* As) and (ii) the *d* metal type (Cu *versus* Ag). As a result, the structural parameters, electronic bands, total and site-projected *l*- decomposed densities of states, and the peculiarities of the inter-atomic interactions for the above Th*MPn*O phases have been obtained and analyzed for the first time.

## 2. Computational aspects

The examined pnictide oxides Th*MPn*O adopt a tetragonal layered structure of the ZrCuSiAs-type, space group *P*4/*nmm*, # 129, Z = 2. The atomic positions are Th: 2*c* (¼,¼,$z_{Th}$); *M*: 2*b* (¾,¼,½); *Pn*: 2*c* (¼,¼,$z_{Pn}$); and O: 2*a* (¾,¼,0); here, $z_{Th}$ and $z_{Pn}$ are the so-called internal coordinates. These structures (see Fig. 1) consist of alternating fluorite-type [$M_2Pn_2$] and anti-fluorite-type [$Th_2O_2$] blocks. In turn, the [$M_2Pn_2$] blocks consist of square nets of *M* atoms, which are tetrahedrally coordinated by four pnictogens. For [$Th_2O_2$] blocks, oxygen atoms are tetrahedrally coordinated by four thorium atoms, and Th is coordinated by four O to form square pyramids.

The calculations of all mentioned Th*MPn*O phases were carried out by means of the full-potential method with mixed basis APW+lo (LAPW) implemented in the WIEN2k suite of programs [13]. The generalized gradient correction (GGA) to exchange-correlation potential in PBE form [14] was used. The *muffin-tin* spheres radii were chosen 2.5 a.u. for Th, 1.6 a.u. for O, and 2.0 a.u. both for Cu/Ag and P/As. The starting configurations were: [Rn]($6d^2 7s^2 7p^0 5f^0$) for thorium, [Ne]($3s^2 3p^3$) for P, [Ar]($3d^{10} 4s^2 4p^3$) for As, [Ar]($3d^{10} 4s^1$) for Cu, [Kr]($4d^{10} 5s^1$) for Ag, and [He]($2s^2 2p^4$) for oxygen. The maximal value for partial waves used inside atomic spheres was $l = 12$ and the maximal value for partial waves used in the computation of *muffin-tin* matrix elements was $l = 4$. The plane-wave expansion with $R_{MT} \times K_{MAX}$ was equal to 7, and $k$ sampling with a 10×10×10 $k$-points mesh in the Brillouin zone was used. Relativistic effects were taken into account within the scalar-relativistic treatment including spin-orbit coupling (SOC). The self-consistent calculations were considered to converge when the difference in the total energy of the crystal did not exceed 0.001 mRy and the difference in the atomic forces did not exceed 1 mRy/a.u. as calculated at consecutive steps.

The hybridization effects were analyzed using the densities of states (DOSs), which were obtained by a modified tetrahedron method [15]. The ionic bonding was considered using Bader [16] analysis. In this approach, each atom of a crystal is surrounded by an effective surface that runs through minima of the charge density, and the total charge of an atom (the so-called Bader charge, $Q^B$) is determined by integration within this region. In addition, some peculiarities of the inter-atomic bonding picture were visualized.

## 3. Results and discussion

*3.1 Structural properties*

As the initial step, the structure of the Th-containing 1111 phases was fully optimized and the equilibrium values of lattice constants and internal coordinates were found. The obtained results are presented in Table 1 in comparison with available experimental data.

It is seen that both *a* and *c* parameters increase as going from Th*M*PO to Th*M*AsO and from ThCu*Pn*O to ThAg*Pn*O. This result can be easily explained by considering the atomic radii of *M* and *Pn*: R(P) = 1.30 Á < R(As) =1.48 Á and R(Cu)=1.28 Á < R(Ag) =1.44 Á.

On the other hand, in the P → As or Cu→ Ag replacements the effect of *anisotropic deformation* of the crystal structure was found. Indeed, as going from ThCuPO to ThCuAsO, the *a* and *c* parameters increase 1.019 and 1.016 times respectively, whereas as going from ThCuPO to ThAgPO, they increase 1.022 and 1.060 times, respectively. The *anisotropic deformation* of the crystal structure owing to atomic substitutions was also found for a set of related layered phases, see for example [2,19,20]. Possibly, this effect (related to strong *anisotropy of inter-atomic bonds*, see also below) has a universal character for all similar layered phases.

*3.2 Electronic properties*

Figures 2-4 and 6 show the band structures and total and atomic-resolved *l*-projected densities of states (DOSs) in ThCuPO, ThCuAsO, ThAgPO, and ThAgAsO phases as calculated for equilibrium geometries.

First of all, since the effect of spin-orbit coupling (SOC) on the electronic structure of materials with light actinides is widely discussed now [21-25], we will compare the results for the band structure and DOSs as obtained in our FLAPW calculations without SOC (non-SOC) and within spin-orbit coupling (using ThCuPO as an example), see Fig. 2.

It is seen that SOC results mainly in a shift and splitting of semi-core Th 6*p* states in the region from -22 eV to -14 eV below the $E_F$. Besides, small splitting of valence bands takes place in the region from -3 eV to -1 eV. However, the common picture of valence bands (and also of DOSs distributions) for ThCuPO as obtained in our calculations without SOC and within spin-orbit coupling varies insignificantly. Further we present the results obtained within SOC.

From Figs. 3 and 4 it is seen that all of these phases will be metallic-like, but with very low densities of states at the Fermi level: $N(E_F)$~0.05-0.17 states/(eV·cell). Therefore these Th-based 1111 phases may be classified as *semimetals* and may be viewed as "intermediate" systems between the group of the above mentioned superconducting 1111 phases and the group of ZrCuSiAs-like semiconductors such as *LnZnPn*O, see [17,26].

Naturally, the topology of the Fermi surface (FS) for Th*MPn*O phases should differ completely from the 2D-like FSs for Fe*Pn*-based 1111 superconducting materials, which consist of electron cylinders around the tetragonal *M* point, and hole cylinders around the *Γ* point, see [2-7]. Indeed, for example for ThAgAsO and ThAgPO, two bands intersect the Fermi level in the vicinity of the *Γ* point and in the *Γ-Z* direction (Fig. 3) forming a Fermi surface, which consists of two isolated closed hole-like and electronic-like pockets around the *Γ* and Z points, respectively, Fig. 5. For ThCuAsO and ThCuPO the Fermi surfaces around *Γ* point are of more complicated character, consisting of enclosed each other two disconnected sheets, Fig. 5.

Let us discuss the common features of the electronic structure of the examined Th-based 1111 phases. Their valence spectra contain three main occupied bands, labeled in Fig. 4 as peaks A, B, and C. The lowest band (peak A lying in the region from about -12 eV to -10 eV) arises mainly from *Pn s* states and is separated from the near-Fermi bands by a gap. These bands (peaks B and C) are located in the energy range from - 7 eV to $E_F$ and are of a mixed type. Indeed, as can be seen from Fig. 6, where the site-projected *l*-decomposed densities of states for ThCuPO are depicted, the peak B contains hybridized Cu 3$d$, O 2$p$, P 3$p$, and Th ($p,d,f$) states, which are responsible for the formation of the covalent component of bonds Cu-P (inside [Cu$_2$P$_2$] blocks) and Th-O (inside [Th$_2$O$_2$] blocks). The insert in Fig. 6 shows also that Th ($p,d,f$) states participate actively in the formation of the valence band, and in its topmost part the contributions from Th 6$d$ and Th 5$f$ states are comparable. Thus, like in metallic thorium [27,28] and a series of thorium compounds with light $sp$ atoms (H, B, C, N,O) [29-34], the Th 5$f$ states for 1111 phases are itinerant and partially occupied. This fact means that the charge state of thorium atoms in 1111-phases differs from that of the purely ionic Th$^{4+}$, see also below. The bottom of the empty conductivity band for the examined phases is formed mostly by the Th 6$d$ and Th 5$f$ states; the contributions of other atoms are negligible.

Quantitative differences between the examined Th*MPn*O phases may be seen from the data presented in Figs. 3, 4 and Table 2, where the band structure parameters (bandwidths) of these materials are summarized as dependent on P → As or Cu → Ag replacements.

*3.3 Inter-atomic bonding*

Let us discuss the bonding picture in the examined Th*MPn*O phases. To describe the **ionic bonding** for these materials, we will start with a purely ionic picture with usual oxidation numbers of atoms: Th$^{4+}$, *M*$^{1+}$, O$^{2-}$, and *Pn*$^{3-}$. So, the charge states of the blocks are [Th$_2$O$_2$]$^{4+}$ and [*M*$_2$*Pn*$_2$]$^{4-}$, *i.e.* the charge transfer (4$e$) occurs from [Th$_2$O$_2$] to [*M*$_2$*Pn*$_2$] blocks. Besides, inside [Th$_2$O$_2$] and [*M*$_2$*Pn*$_2$] blocks, the ionic bonding takes place between ions with opposite charges: Th - O and *M* - *Pn*.

To estimate numerically the amount of electrons redistributed between various atoms and between adjacent [Th$_2$O$_2$]$^{\delta+}$/[*M*$_2$*Pn*$_2$]$^{\delta-}$ blocks, we carried out a Bader [16] analysis, and the obtained effective atomic charges ΔQ are presented in Table 3. These results show that the inter-atomic and inter-blocks charge transfer is considerably smaller than it was predicted in the idealized ionic model. Namely, the transfer $\Delta Q([Th_2O_2] \to [M_2Pn_2])$ is about 2.08-1.94 $e$; these values are smaller for Cu-containing phases than for their Ag-containing counterparts and are smaller for arsenides as compared with the related phosphides. These results confirm that all Th*MPn*O phases are partially ionic compounds – owing to the presence of appreciable covalent contributions.

Indeed, the character of **covalent Th-O and M-Pn bonding** in Th*MPn*O phases (owing to hybridization of the states Th ($p,d,f$)- O 2$p$ and $M$ $d$ – $Pn$ $p$, respectively) *inside* [$Th_2O_2$] and [$M_2Pn_2$] blocks may be understood from site-projected DOSs calculations, see above. These bonds are also well visible on the 3D charge density distribution picture, see Fig. 1. On the other hand, no direct inter-atomic bonds **between** the adjacent blocks are present.

Thus, summarizing the above results, the picture of chemical bonding for Th*MPn*O phases may be described in the following way. Inside [$Th_2O_2$] blocks mixed covalent-ionic bonds Th-O take place (owing to hybridization of Th ($p,d,f$)- O 2$p$ states and Th → O charge transfer), while the estimated effective charges of Cu and Ag atoms in 1111-compounds are small. Therefoer it should be expected that the covalent component of *M-Pn* bonding is predominant *inside* [$M_2Pn_2$] blocks. Between the adjacent [$Th_2O_2$]/[$M_2Pn_2$] blocks, ionic bonds emerge owing to [$Th_2O_2$] → [$M_2Pn_2$] charge transfer. Generally, the bonding in Th*MPn*O phases can be classified as a high-anisotropic mixture of ionic and covalent contributions.

## 4. Conclusions

In summary, by means of the FLAPW-GGA approach, we studied the structural, electronic, and chemical bonding picture for four ZrCuSiAs-like Th-based quaternary pnictide oxides ThCuPO, ThCuAsO, ThAgPO, and ThAgAsO.

Our results show that the replacements of $d$ metal atoms (Cu ↔ Ag) and pnictogen atoms (P ↔ As) lead to *anisotropic deformations* of the crystal structure; this effect is related to strong anisotropy of inter-atomic bonds.

Our studies showed that the examined Th*MPn*O phases may be classified as *semimetals* and may be viewed as "intermediate" systems between the main groups of superconducting and semiconducing ZrCuSiAs-like phases. The Fermi surfaces for these Th-based 1111 phases (consisting of isolated closed hole-like and electronic-like pockets) are completely different from the FSs for Fe*Pn*-based 1111 superconducting materials, which consist of a set of 2D-like electron and hole cylinders. The data obtained show also that Th $5f$ states participate actively in the formation of valence bands. In other words, the Th $5f$ states for 1111 phases are itinerant and partially occupied.

Finally, we found that the bonding in Th*MPn*O phases can be classified as a high-anisotropic mixture of ionic and covalent contributions, where mixed covalent-ionic bonds take place inside [$Th_2O_2$] and [$M_2Pn_2$] blocks, whereas between the adjacent [$Th_2O_2$]/[$M_2Pn_2$] blocks, ionic bonds emerge owing to [$Th_2O_2$] → [$M_2Pn_2$] charge transfer.

**Acknowledgements**
This work was supported by RFBR, grants 09-03-00946 and 10-03-96008.

**Table 1.**
The optimized lattice parameters ($a$ and $c$, in Å) and internal coordinates ($z_{Th}$, $z_{P, As}$) for layered 1111-like Th-based pnictide oxides.

| phase | $a$, Å | $c$, Å | $z_{Th}$ | $z_{P, As}$ |
|---|---|---|---|---|
| ThCuPO | 3.912 (3.894[a]) | 8.238 (8.283[a]) | 0.1715 | 0.3323 |
| ThCuAsO | 3.990 (3.961[a]) | 8.372 (8.440[a]) | 0.1669 | 0.3253 |
| ThAgPO | 3.998 (3.966[a]; 3.961[b]) | 8.731 (8.786[a]; 8.778[b]) | 0.1590 (0.1546[b]) | 0.2988 (0.2979[b]) |
| ThAgAsO | 4.054 | 8.914 | 0.1545 | 0.2979 |

* available experimental data are given in parentheses: [a] Ref. [17], and [b] Ref. [18]

**Table 2.**
Calculated valence band structure parameters (band width, eV) for layered 1111-like Th-based pnictide oxides.

| Phase / band type * | 1 | 2 | 3 | 4 |
|---|---|---|---|---|
| ThCuPO | 11.87 | 1.55 | 3.45 | 6.87 |
| ThCuAsO | 12.17 | 1.36 | 3.97 | 6.84 |
| ThAgPO | 11.11 | 1.25 | 2.88 | 6.98 |
| ThAgAsO | 11.68 | 1.14 | 3.48 | 7.06 |

* band types: 1 – common valence band ($Pn\ s \div E_F$); 2 - $Pn\ s$ band; 3 - band gap: $Pn\ s$ – near-Fermi (Th $p,d,f$ + $M\ d$ + O $2s$ + $Pn\ p$) band; 4 - near-Fermi (Th $p,d,f$ + $M\ d$ + O $2s$ + $Pn\ p$) band;

**Table 3.**
The atomic effective charges ($\Delta Q$, in $e$) and inter-blocks charge transfer ($\Delta Q([Th_2O_2] \rightarrow [M_2Pn_2])$, in $e$) for layered 1111-like Th-based pnictide oxides as estimated within the Bader scheme.

| phase / $\Delta Q$ | Th | $M$ | $Pn$ | O | $\Delta Q([Th_2O_2] \rightarrow [M_2Pn_2])$ |
|---|---|---|---|---|---|
| ThCuPO | 2.33 | 0.11 | -1.14 | -1.32 | -2.02 |
| ThCuAsO | 2.29 | 0.05 | -1.03 | -1.33 | -1.94 |
| ThAgPO | 2.36 | 0.03 | -1.08 | -1.32 | -2.08 |
| ThAgAsO | 2.32 | -0.02 | -0.98 | -1.33 | -1.98 |

**FIGURES**

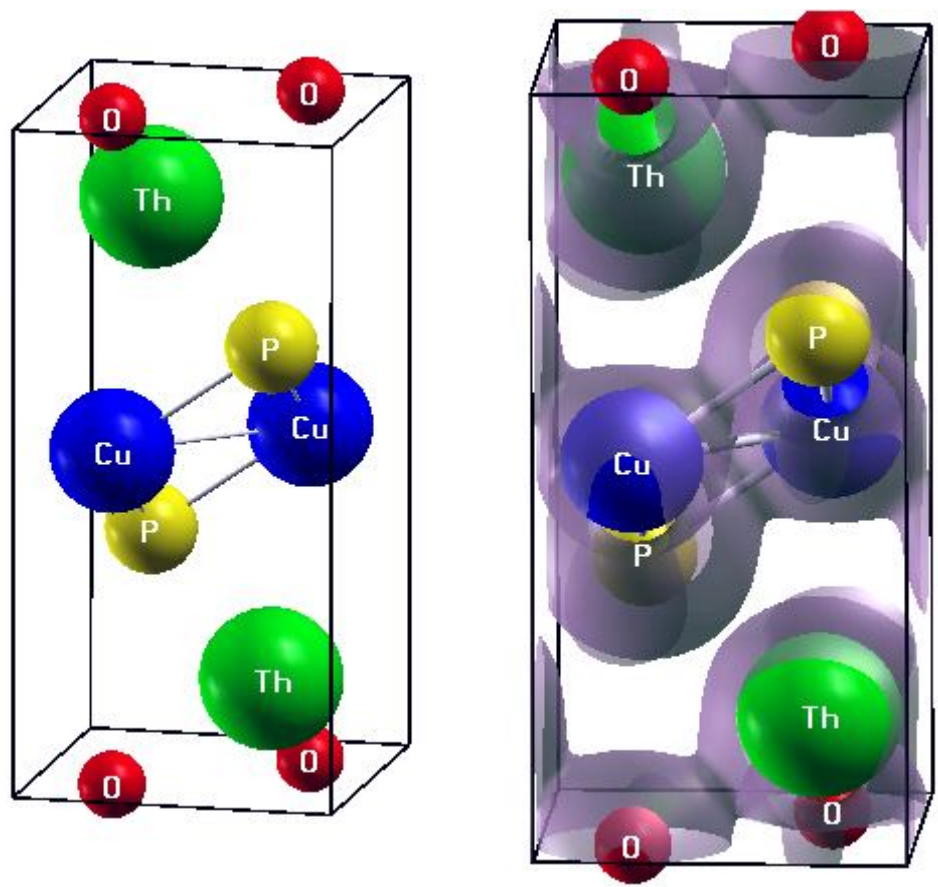

**Fig. 1.** (*Color online*) The crystal structure (*left*) and the valence charge density map (*right*) for ThCuPO. The isosurface corresponds to $\rho=0.05$ e/Å$^3$.

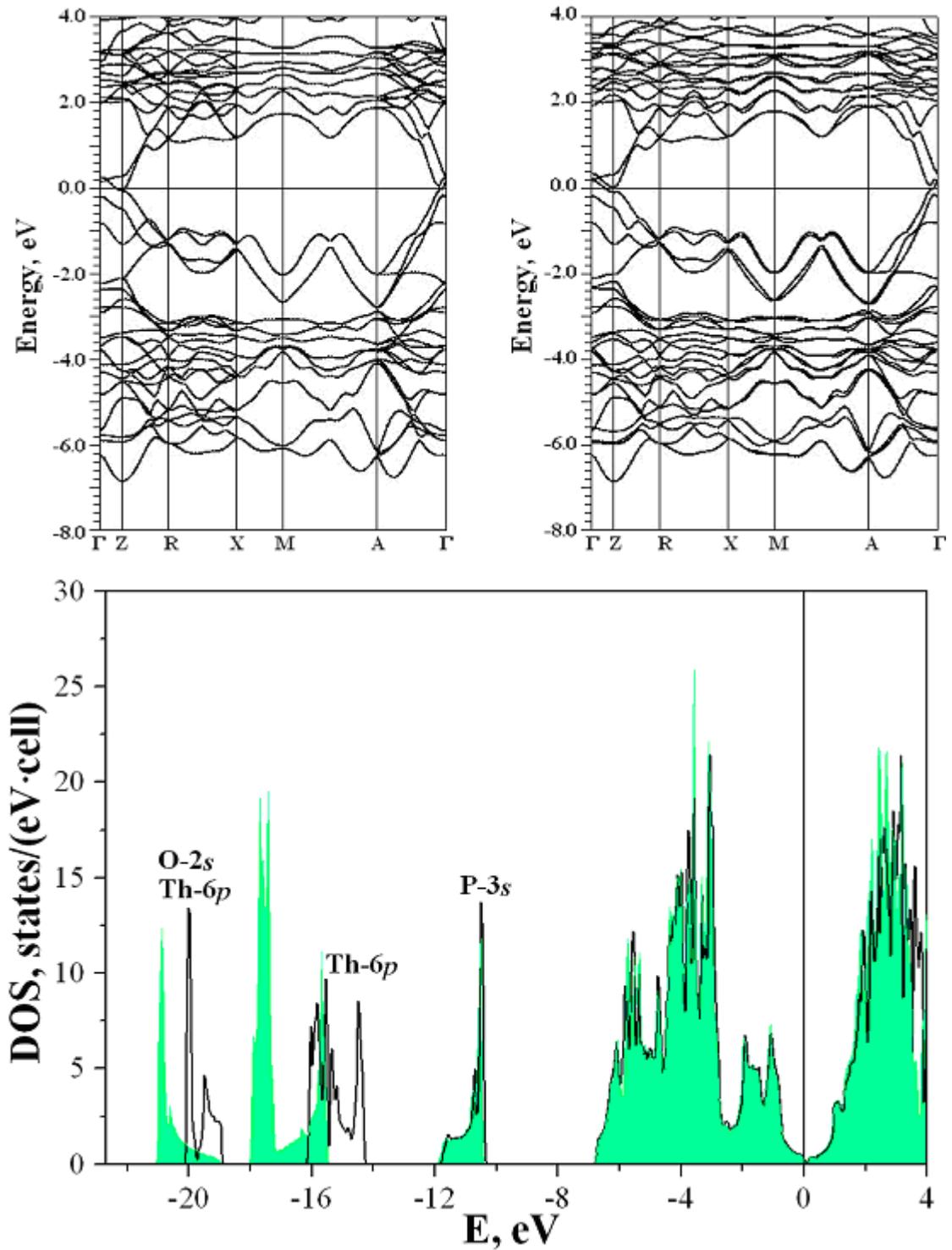

**Fig. 2.** (*Color online*) *Upper panel*: the band structure of ThCuPO calculated without (*left*) and within SOC (*right*). *Lower panel* – total DOSs of ThCuPO calculated without (*filled*) and within SOC (*black line*).

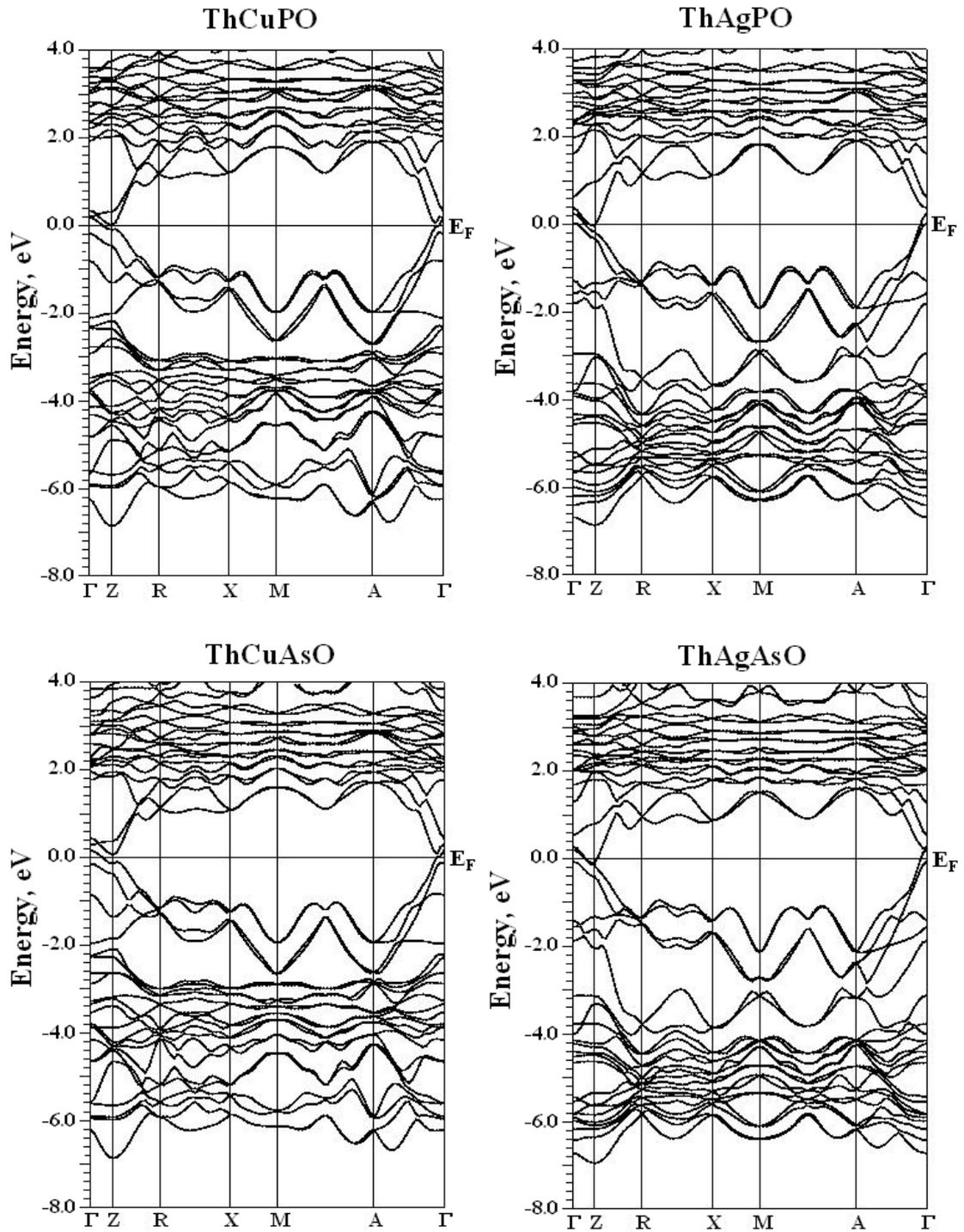

**Fig. 3.** Band structures of ThCuPO, ThAgPO, ThCuAsO, and ThAgAsO.

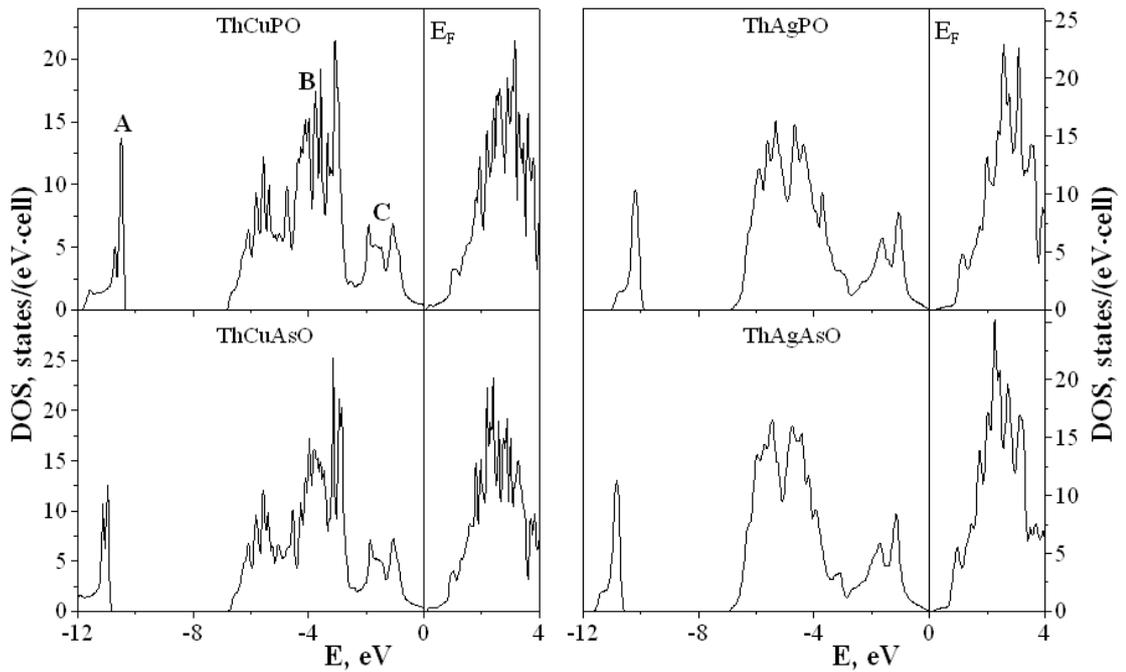

**Fig. 4.** Total densities of electronic states for ThCuPO, ThAgPO, ThCuAsO, and ThAgAsO.

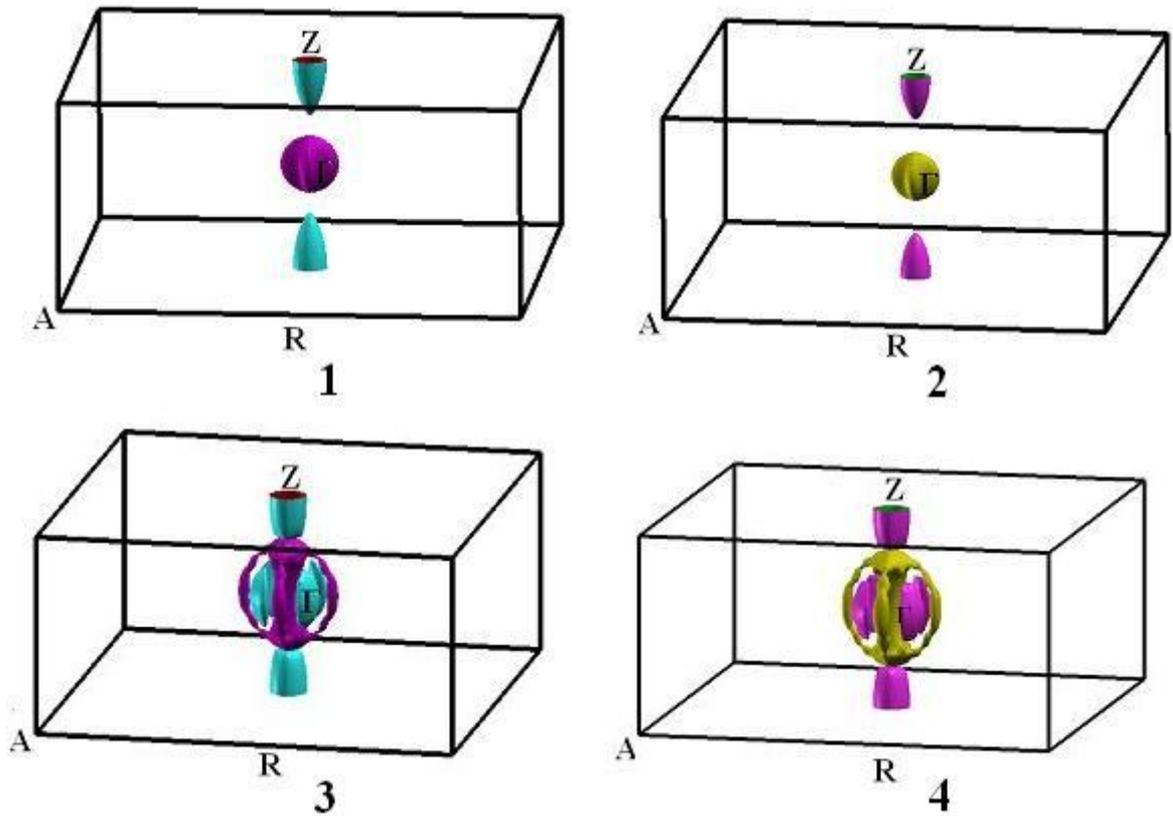

**Fig. 5.** (*Color online*) Fermi surfaces for ThAgAsO (1), ThAgPO (2), ThCuAsO (3), and ThCuPO (4).

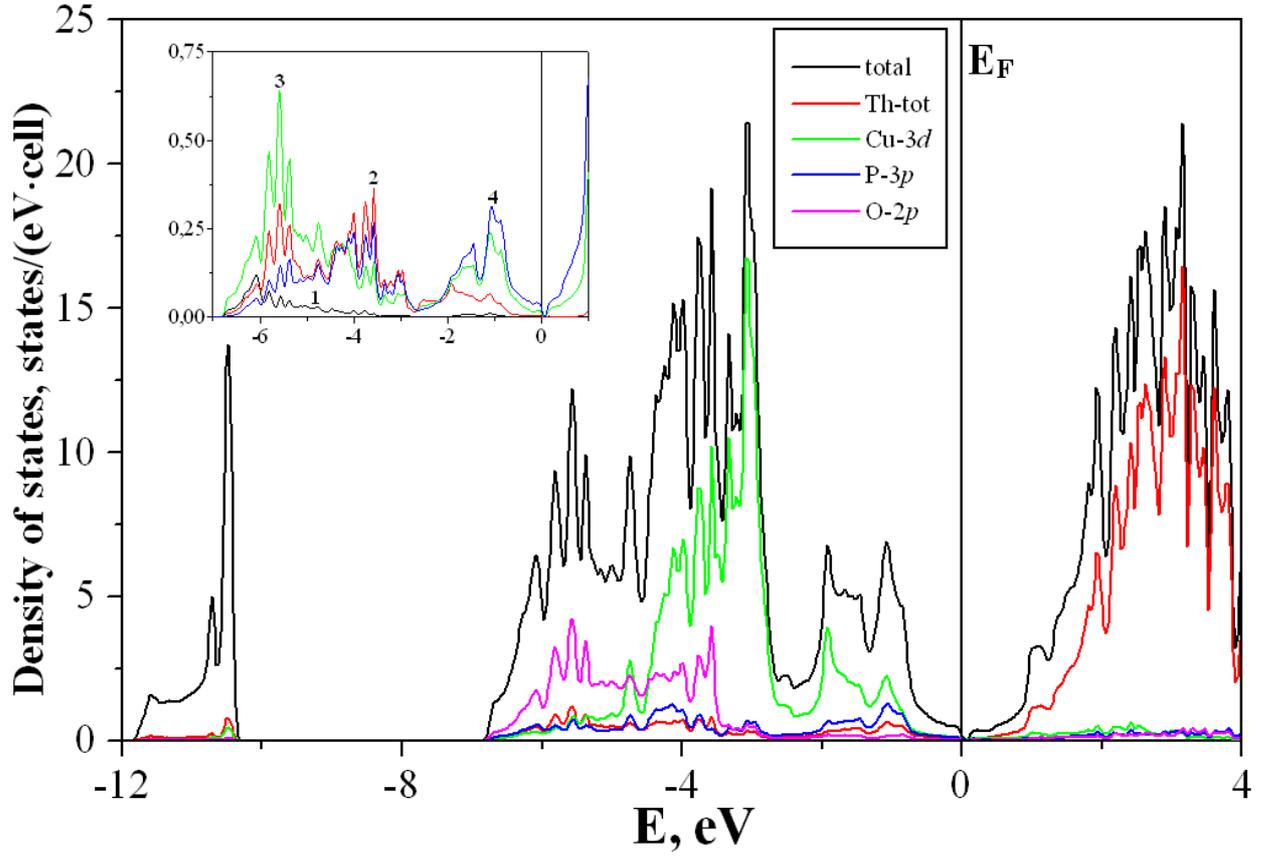

**Fig. 6.** (*Color online*) Total and partial atomic- and *l*-projected densities of states for ThCuPO. *Insert* – the *l*-projected DOS of thorium in the near-Fermi region (1 – Th-*s* states, 2 – Th-*p* states, 3 – Th-6*d* states, and 4 – Th-5*f* states).